\documentclass[5p]{elsarticle}

\usepackage{multirow,url,graphicx,siunitx,textcomp,helvet,array,makecell,hhline,lineno,hyperref,epstopdf}
\newcolumntype{M}[1]{>{\centering\arraybackslash}m{#1}}
\DeclareSIUnit\px{px}
\modulolinenumbers[5]

\journal{Measurement}

\begin{document}

\begin{frontmatter}

\title{Thermometry of intermediate level waste containers using phosphor thermometry and thermal imaging}

\author{J L McMillan\fnref{fn1}}

\author{A Greenen, W Bond, M Hayes, R Simpson, G Sutton and G Machin}
\address{National Physical Laboratory, Hampton Road, Teddington, TW11 0LW, UK}

\author{J Jowsey and A Adamska}
\address{Sellafield Site, Seascale, Cumbria, CA20 1PG}

\fntext[fn1]{Corresponding author \texttt{jamie.mcmillan@npl.co.uk}}

\begin{abstract}
Intermediate level waste containers are used for the storage of an assortment of radioactive waste. This waste is heat-generating and needs monitoring and so this work was undertaken to determine whether the mean {\em internal container temperature can be inferred from the temperature of the vent}. By using two independent thermometry techniques -- phosphor thermometry and thermal imaging -- the internal temperature was demonstrated to be proportional to the vent temperature as measured by both methods. The correlation is linear and given suitable characterisation could provide robust indication of the internal bulk temperature.
\end{abstract}

\begin{keyword}
thermal imaging \sep infrared \sep radiation thermometry \sep thermography \sep temperature measurement \sep metrology \sep  phosphor thermometry \sep nuclear \sep waste container \sep ILW \sep decommissioning
\end{keyword}

\end{frontmatter}



\section{\bf Scope of Work} \label{sec:scope}

Nuclear fission is a major part of the energy infrastructure of the UK. However the decommissioning of nuclear facilities requires the safe and sustainable storage of its spent fuel and other radioactive by-products. One form of this waste, intermediate level waste (ILW), mostly comprises nuclear reactor components, graphite and sludges from the treatment of radioactive liquid effluents \cite{ref:nda_ilw_waste}. Typically ILW is stored in cylindrical steel containers, shown in Figure \ref{fig:drumSchematic}. The container has a pair of dewatering tubes, a sintered gauze above the waste and a meshed vent on the lid.

ILW containers are stored in ventilated engineered stores but a qualitative inspection method to assess the physical properties of each drum (e.g. dimension, mass or temperature) does not currently exist. The objective in this work is to answer whether {\em internal container temperature can be inferred from the temperature of the vent}. 

Two methods were used to determine the vent temperature: thermal imaging and phosphor thermometry. Both are non-contact methods, reducing the complexity for any future integration, as well as being instantaneous and dynamic. The temperatures obtained by the two methods were compared against those determined from the internal contact thermometers.  

It is currently not understood whether the phosphor thermometry method proposed in this work would be suitable with intermediate or high level waste, due to the gamma radiation flux influencing measurement. This will be addressed in future studies.

This work was performed at the National Physical Laboratory (NPL). The temperature group at NPL houses the realisation of the International Temperature Scale of 1990 (ITS-90) for the UK and has a wealth of experience in demonstrating traceability to ITS-90 in a range of challenging environments \cite{ref:npl_its90}. All sensors in this work were calibrated to ITS-90, in addition end-to-end uncertainty analysis was undertaken for all measurements to ensure equivalence of measurment methods could be robustly demonstrated.


\section{\bf Introduction to the Measurement Methods} \label{sec:introduction}
\subsection{Thermal imaging} \label{subsec:introduction_thermal_imagers}

The physics of thermal imagers is an extension of radiation thermometry, which is based on determining the spectral intensity \(I\) emitted by a surface and which is described by the Planck distribution law \cite{ref:radiation_thermometry}. The signal measured by each pixel of a thermal imager is proportional to the intensity emitted by the surface being measured.

\begin{equation}
I(\lambda,T,\theta,\phi) = \frac{\varepsilon(\lambda,T,\theta,\phi) c_1}{\lambda^5 \left[ exp(c_2 / \lambda T) - 1 \right]}
\label{eq:planck}
\end{equation}

Spectral intensity is a function of the surface emissivity \(\varepsilon\), wavelength \(\lambda\), temperature of the surface \(T\) and the first and second radiation constants \(c_1\) and \(c_2\). The emissivity of a surface describes how efficiently that surface can absorb and emit thermal radiation. Emissivity is itself dependent on the wavelength, the temperature, and both polar and azimuthal angles of measurement \((\theta,\phi)\). 

Emissivity is typically the largest contributor of uncertainty to radiation thermometry due to the sparcity of data, its variability with material, surface finish, temperature dependence and spectral distribution.

\begin{figure}
\centering
\includegraphics[width=0.5\textwidth,keepaspectratio]{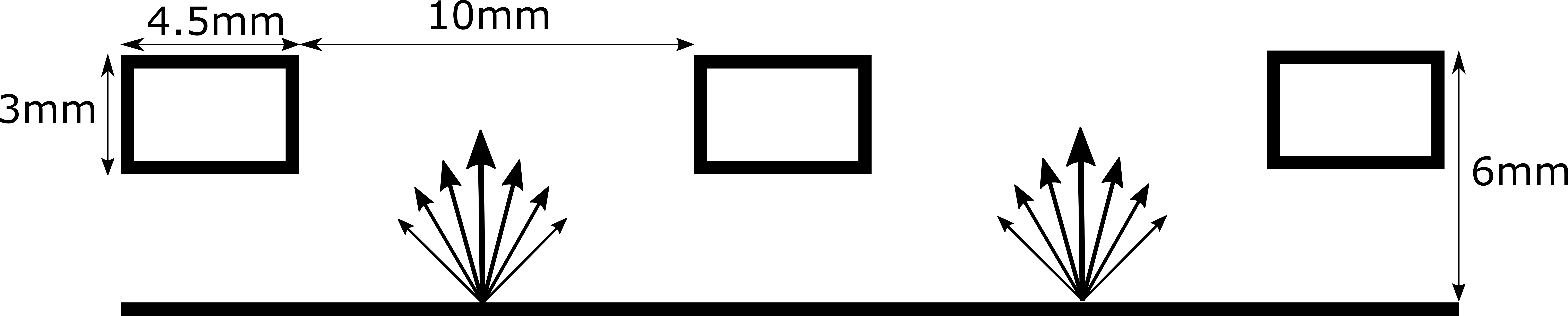}
\caption{A schematic detailing the vent arrangement on the container lid. This depicts the geometrical enhancement of emissivity.}    
\label{fig:blackbody_vent}
\end{figure}

Part of this investigation was to assess whether the geometry of the vent at the top of the ILW drum enhanced the emissivity, enabling more robust non-contact temperature measurement. It is known that geometric enhacement of emissivity exists due to surface features such as holes and wedges \cite{ref:emissivity_of_cavities,ref:npl_reference_blackbody}. By making use of multiple reflections and surface cavities, the emissivity can be increased to reduce temperature errors, a fact taken advantage of as below (see Figure~\ref{fig:blackbody_vent}). The emissivity of a surface can also be increased through the application of a high emissivity coating. 

Due to the unknown angular variation of emissivity for a surface, and whether this is specular or diffuse, the angular variation of spectral intensity is unknown. If the surface is particularly specular, it is known that at angles far from normal, the emissivity will be lower and hence the measured temperature will be lower than at normal angles \cite{ref:brdf}. Both this effect and the geometrical enhancement would in this case compete to potentially normalise the angular variation of temperature across the vent surface. 

\subsection{Phosphor thermometry} \label{subsec:introduction_phosphor}

\begin{figure}[ht]
\centering
\includegraphics[width=0.45\textwidth,keepaspectratio]{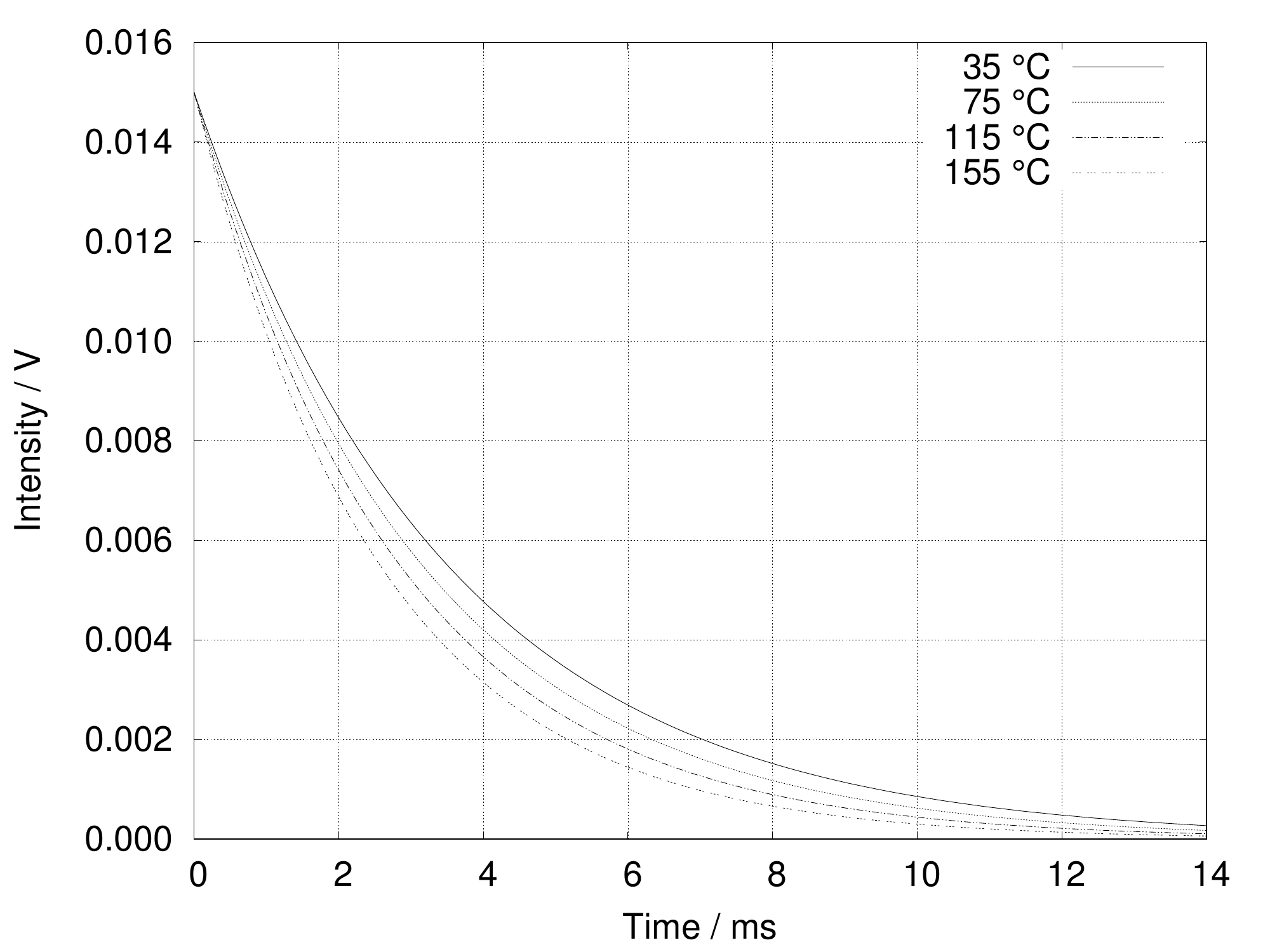}
\caption{Typical decay curves for the phosphor-binder combination used throughout this work.}
\label{fig:phosphor_decay}
\end{figure}

Phosphor thermometry provides a method to remotely determine the temperature of a surface that is independent of both surface emissivity and background thermal radiation. The technique requires a thin coating of thermographic phosphor mixed with a binder (less than \SI{150}{\micro\metre} in thickness) to be sprayed on the surface, this is then interrogated optically. The phosphor used within this study was manganese-activated mangnesium fluorogerminate Mg\textsubscript{4}FGeO\textsubscript{6}:Mn. The phosphor used was excited with violet light (\SI{420}{\nano\metre}) and, once the excitation was removed, the phosphorescence decay at \SI{660}{\nano\metre} was measured and the decay time, which can be directly related to temperature through calibration, determined as 

\begin{equation}
I(t) = I_0\exp^{-t/\tau}
\label{eq:phosphor_decay}
\end{equation}

\noindent and the decay time \(\tau\) is determined from a least squares fit of Equation~\ref{eq:phosphor_decay} to the measurement data. Where \(I(t)\) is the intensity at a time \(t\) and \(I_0\) is the initial intensity. Figure~\ref{fig:phosphor_decay} shows typical decay curves for the phosphor used in this investigation from \SIrange{35}{115}{\celsius}. For the ILW measurements, a spot measurement system was used \cite{ref:empress}.

To determine the surface temperature from the measured decay time a calibration is required: this was performed as follows. A thin phosphor coating was applied to a stainless steel calibration target and the decay time versus temperature  was determined over the temperature range \SIrange{20}{525}{\celsius}. The target incorporated three calibrated (type N, \SI{1}{\milli\metre} diameter) mineral insulated thermocouples placed at depths of \SI{2}{\milli\metre}, \SI{7}{\milli\metre} and \SI{12}{\milli\metre} below the surface such that their tips were located centrally and aligned axially within the target. The true surface temperature was determined by linear extrapolation of these three measurements. 


\section{\bf Measurements} \label{sec:measurements}
\begin{figure}[h]
\centering
\includegraphics[width=0.5\textwidth,keepaspectratio]{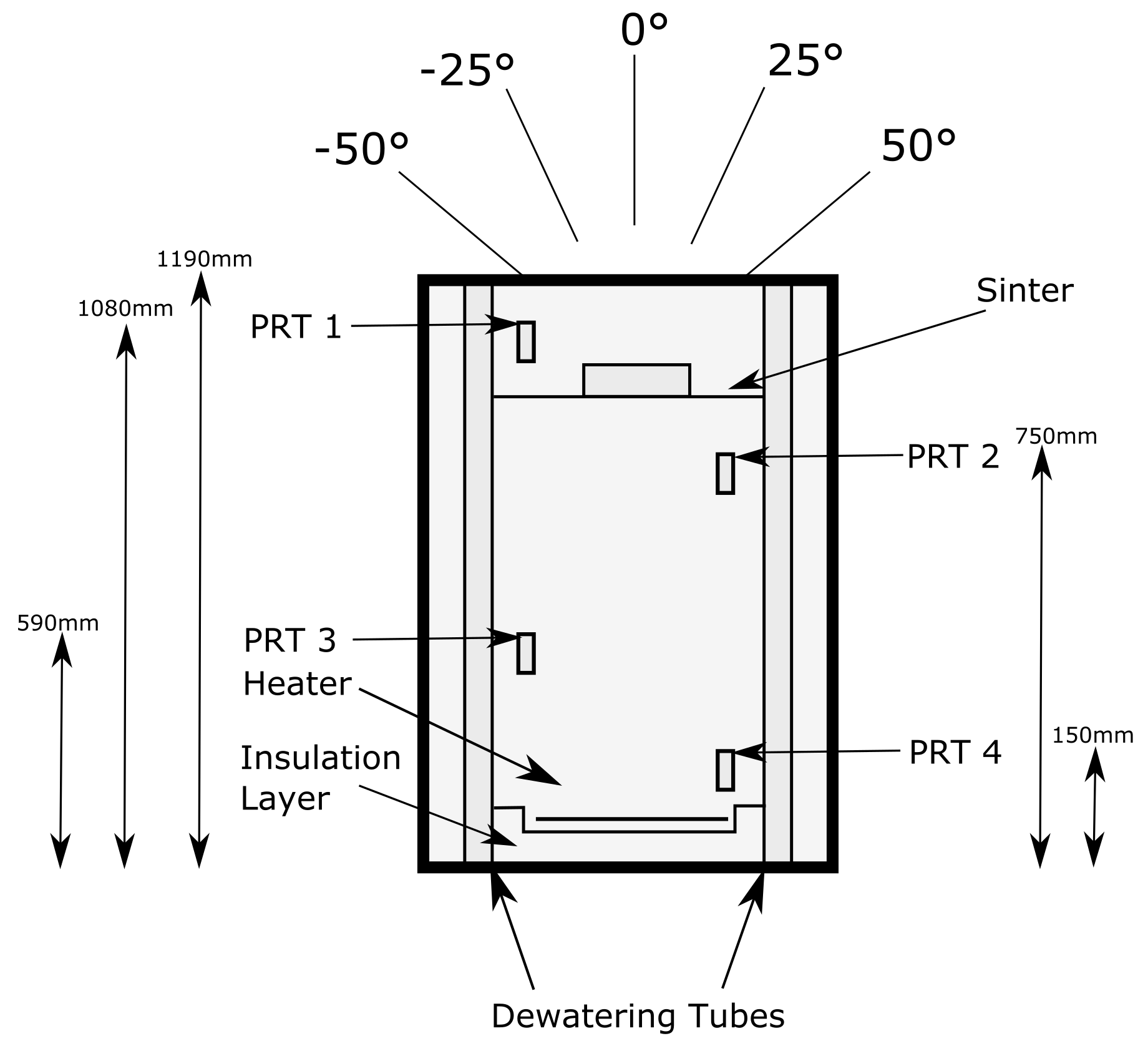}
\caption{A schematic showing the internal instrumentation of the drum. Distances were measured from the bottom of the drum to the top of the PRT. The angles of observation depict the positions used for the thermal imagers.}
\label{fig:drumSchematic}
\end{figure}

A schematic of the ILW container (hereafter referred to as the drum) is seen in Figure~\ref{fig:drumSchematic}. In order to supply heat to the drum, two heaters were placed at the bottom of the drum. The heaters were each connected to a benchtop controller that regulated the power to the heaters, in order to obtain a stable temperature setpoint.

Four class A thin film Pt100 platinum resistance thermometers (PRTs) \cite{ref:iso751} were each potted in a \SI{40}{\milli\metre} long cylinder (\SI{4}{\milli\metre} in diameter). These PRTs were fixed to the internal dewatering tubes (seen in Figure~\ref{fig:drumSchematic}) within the drum to provide a measurement of the internal bulk temperature. A compressed air line was passed into the drum to provide a comparable environment to in-storage drums. Typical flow rates during measurement were nominally less than \SI{0.4}{\liter\per\minute}.

For this investigation two thermal imaging systems were employed: the long-wave infrared (LWIR) (\SIrange{7.5}{13.5}{\micro\metre}) FLIR Tau 2 microbolometer and the medium-wave infrared (MWIR) (\SIrange{2.0}{5.7}{\micro\metre}) InfraTec ImageIR 8300 indium antimonide photon detector. Both were equivalent in array size (\(640 \times 512~\)\SI{}{\px}) but exhibit different thermal sensitivies and form factors.

In order to measure any angular dependency of the temperature measured from a thermal imager, the imagers had to be mounted so that they could be positioned at a number of angles about the vent of the container. The LWIR imager was mounted on an articulated optical post for the off-normal angles, and for the perpendicular position a simple arm was used; the angles of incidence can be seen in Figure~\ref{fig:drumSchematic}. For the MWIR measurements, a different setup was used and measurements at only two observation angles were performed.

\begin{figure}[h]
\centering
\includegraphics[width=0.45\textwidth,keepaspectratio]{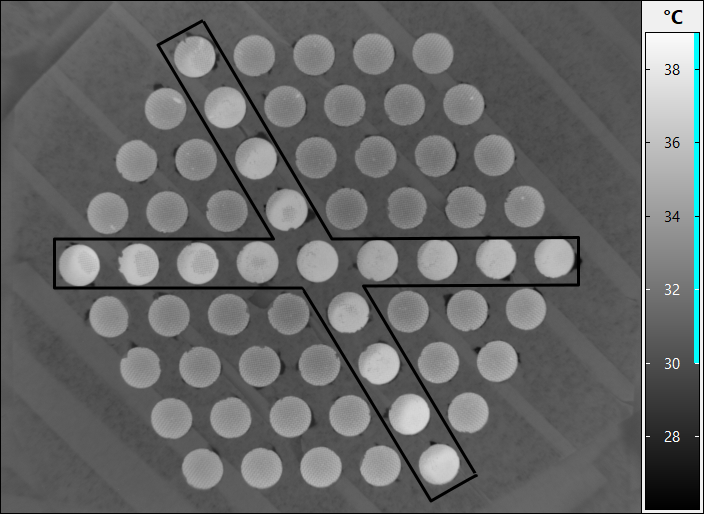}
\caption{Typical example thermal image obtained during the measurement campaign. This was from the MWIR thermal imager and the drum set to the highest setpoint. Along two of the central axes (highlighted), the whiter regions indicate the vents coated with phosphor. The phosphor has a higher emissivity than the measured surface, hence appears brighter.}
\label{fig:exp_thermal_image}
\end{figure}

The phosphor thermometry system was originally designed to be used as a hand-held surface probe. For the experiments in this study, additional lenses were used which allowed remote operation of the probe to bring the phosphor system out of the field of view of the thermal imager. For the measurement of the vent, a coating of phosphor was applied in a cross-like arrangement across the individual vents, as shown in Figure~\ref{fig:exp_thermal_image}. 

A typical image measured by a thermal imager can be seen in Figure~\ref{fig:exp_thermal_image}. The region coated by the phosphor is clearly depicted by the whiter holes. To differentiate between the two emissivity regimes, the coated holes were defined as the high emissivity vents and the remaining were known as the low emissivity vents. During data analysis these two regimes were assessed separately; for each image (at each setpoint and angle) a spot was manually placed at the centre of each vent. This spot then defined a \(3 \times 3\) grid of pixels whose indicated temperature was averaged; the average and standard deviation of all high emissivity and low emissivity vents were recorded. 

It should be noted that the temperatures measured by a thermal imager have not been corrected for the emissivity of the surface measured, therefore these temperatures should be regarded as apparent temperatures. For all thermal imager temperature measurements the emissivity of the imager was set to \num{1.00}.

\section{\bf Results} \label{sec:results}

\begin{figure}[h]
\centering
\includegraphics[width=0.45\textwidth,keepaspectratio]{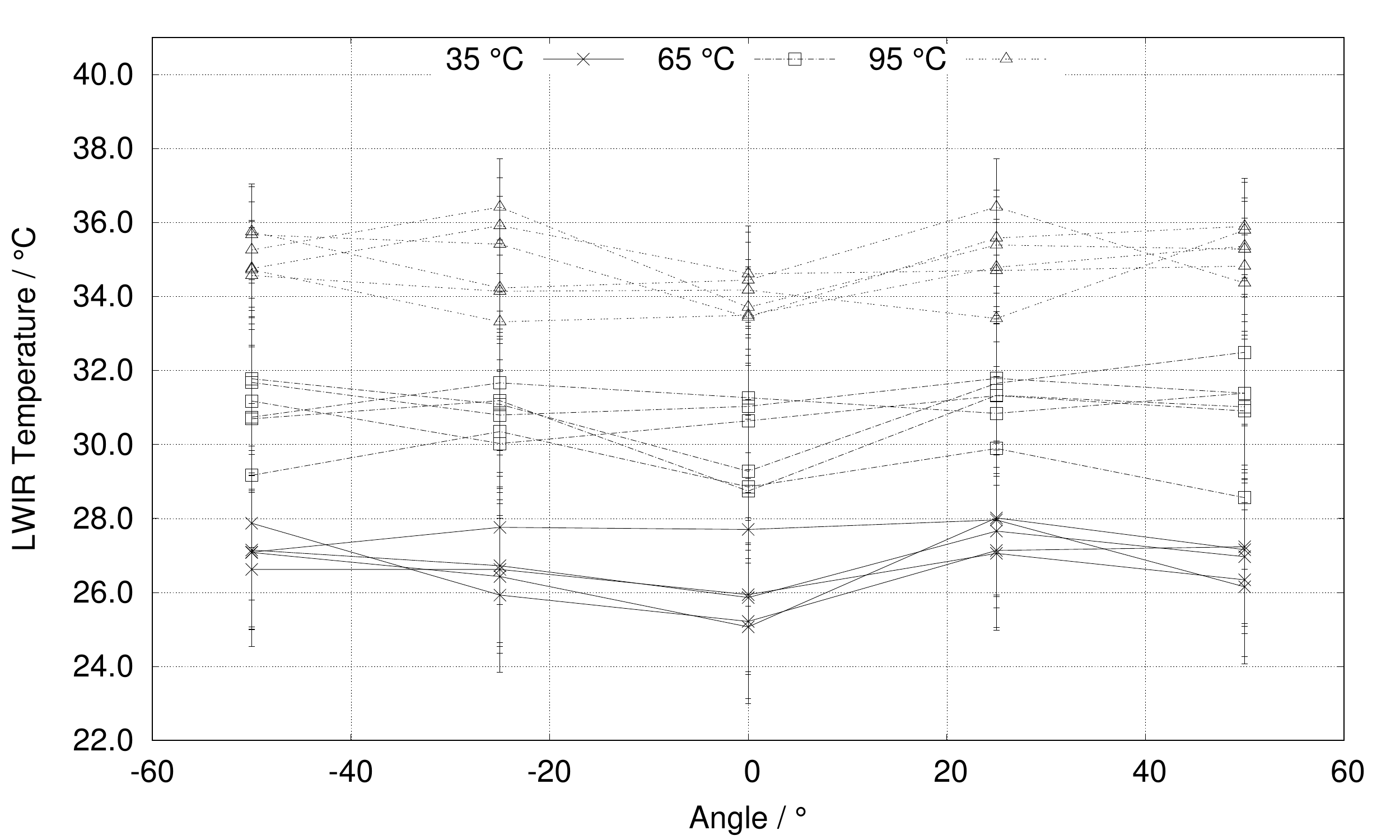}
\caption{Temperature measured for the low emissivity vents by the LWIR thermal imager at each temperature setpoint across a number of angles and through multiple measurement sets. \SI{0}{\degree} corresponds to the imager perpendicular to the drum vents. The three data sets labelled correspond to the bottom, middle and top temperature setpoints from the benchtop controller.}
\label{fig:results_tau2_angular_low}
\end{figure}

\begin{figure}[h]
\centering
\includegraphics[width=0.45\textwidth,keepaspectratio]{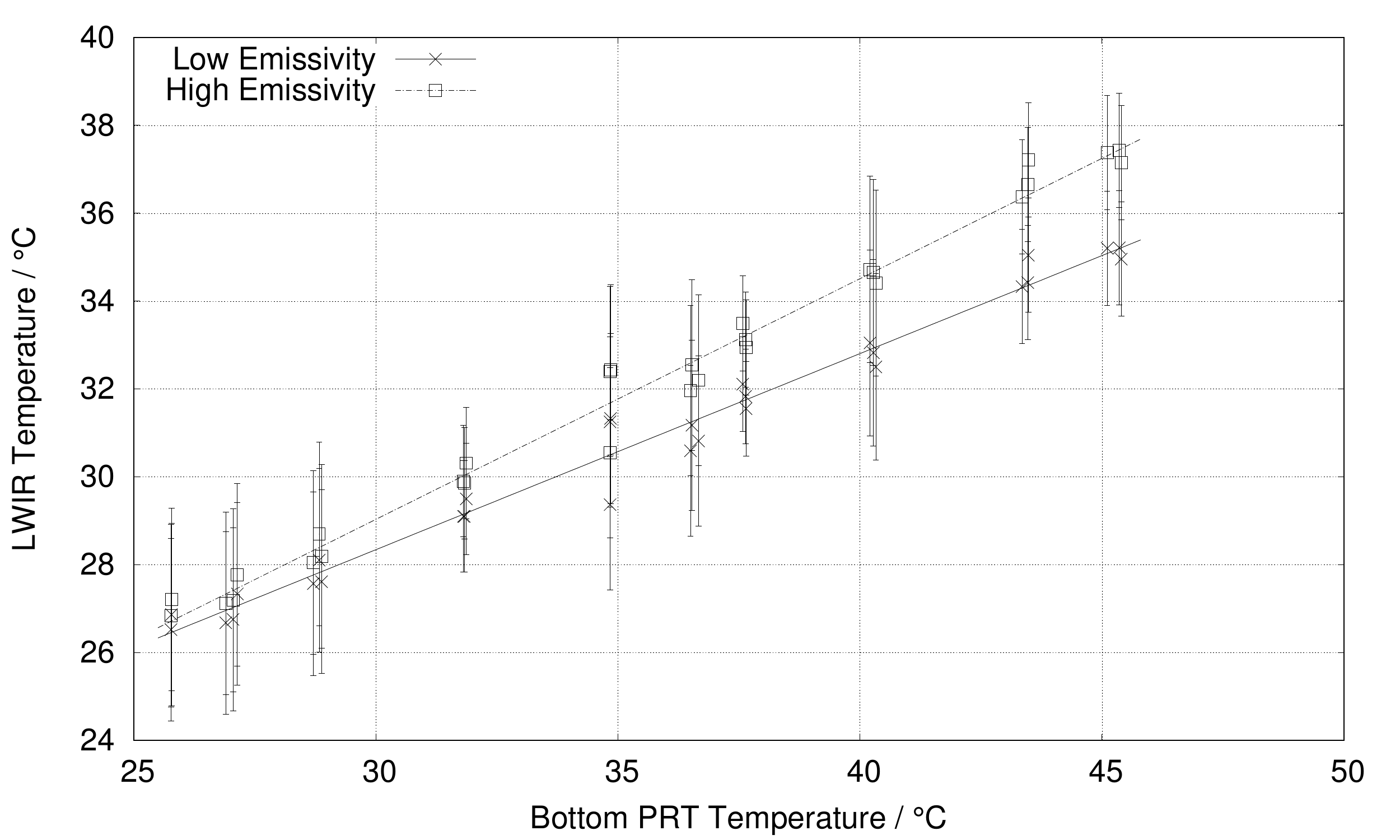}
\caption{The temperature measured by the LWIR thermal imager at each setpoint, plotted against the temperature measured by the bottom-most PRT in the ILW container. The temperatures at each angle and data set from Figure~\ref{fig:results_tau2_angular_low} was averaged for each point in this plot; the error bars indicate the uncertainty from Section \ref{sec:uncertainty}.}
\label{fig:results_tau2_average}
\end{figure}

This section describes the results of the thermal imaging and phosphor thermometry. The measurements shown in Figure~\ref{fig:results_tau2_angular_low} depict the temperatures measured by the LWIR thermal imager at each setpoint through a number of different angles of incidence. The data shown in Figure~\ref{fig:results_tau2_average} is the average of the multiple data sets for each angle and temperature setpoint The uncertainty for each value is discussed in Section \ref{sec:uncertainty}.

\begin{figure}[h]
\centering
\includegraphics[width=0.45\textwidth,keepaspectratio]{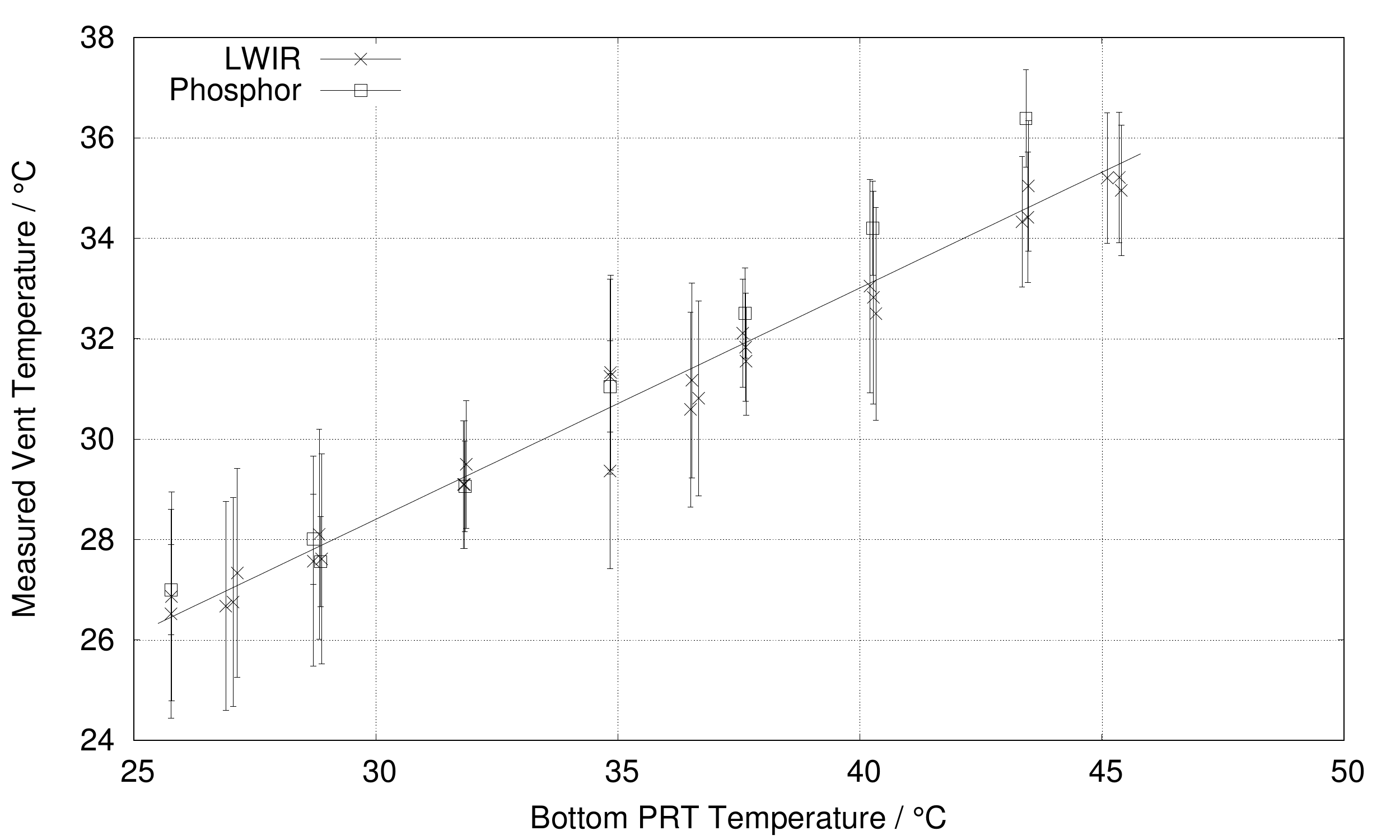}
\caption{Comparison between phosphor and thermal imaging data. The thermal imaging results were measured by the LWIR imager and show the low emissivity vents. The line depicts the best fit of measurement between the two systems.}
\label{fig:results_phosphor_vs_thermal_imaging}
\end{figure}

The data in Figure~\ref{fig:results_phosphor_vs_thermal_imaging} shows both the phosphor and LWIR thermal imager temperatures compared to the temperature measured by the bottom-most PRT (PRT 4). Each measurement is an average across each angle and over nominally a twenty minute period.

\begin{figure}[h]
\centering
\includegraphics[width=0.45\textwidth,keepaspectratio]{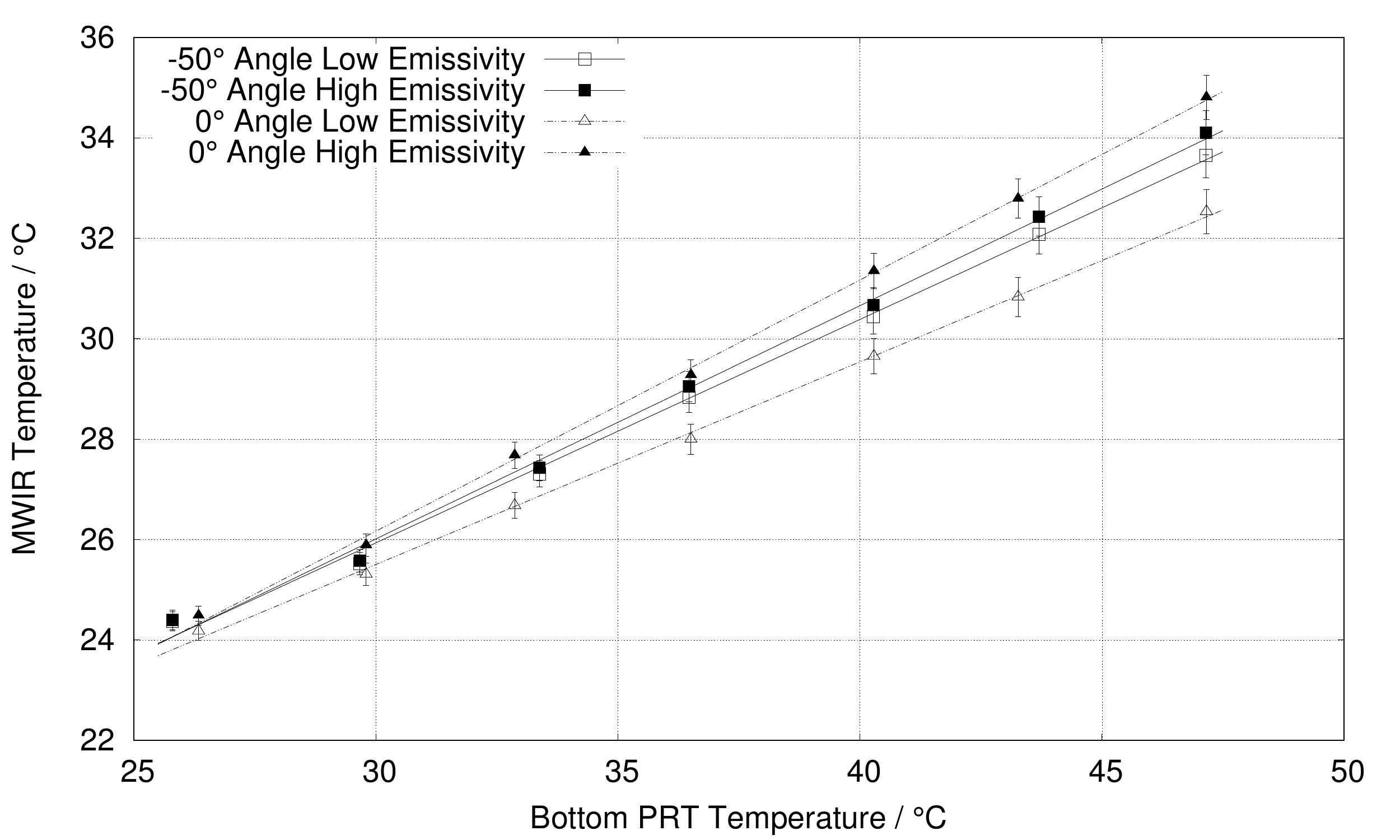}
\caption{The temperature measured by the MWIR thermal imager at each setpoint, plotted against the temperature measured by the bottom-most PRT in the ILW container.}
\label{fig:results_thermal_imaging_infratec}
\end{figure}

\begin{figure}[h]
\centering
\includegraphics[width=0.45\textwidth,keepaspectratio]{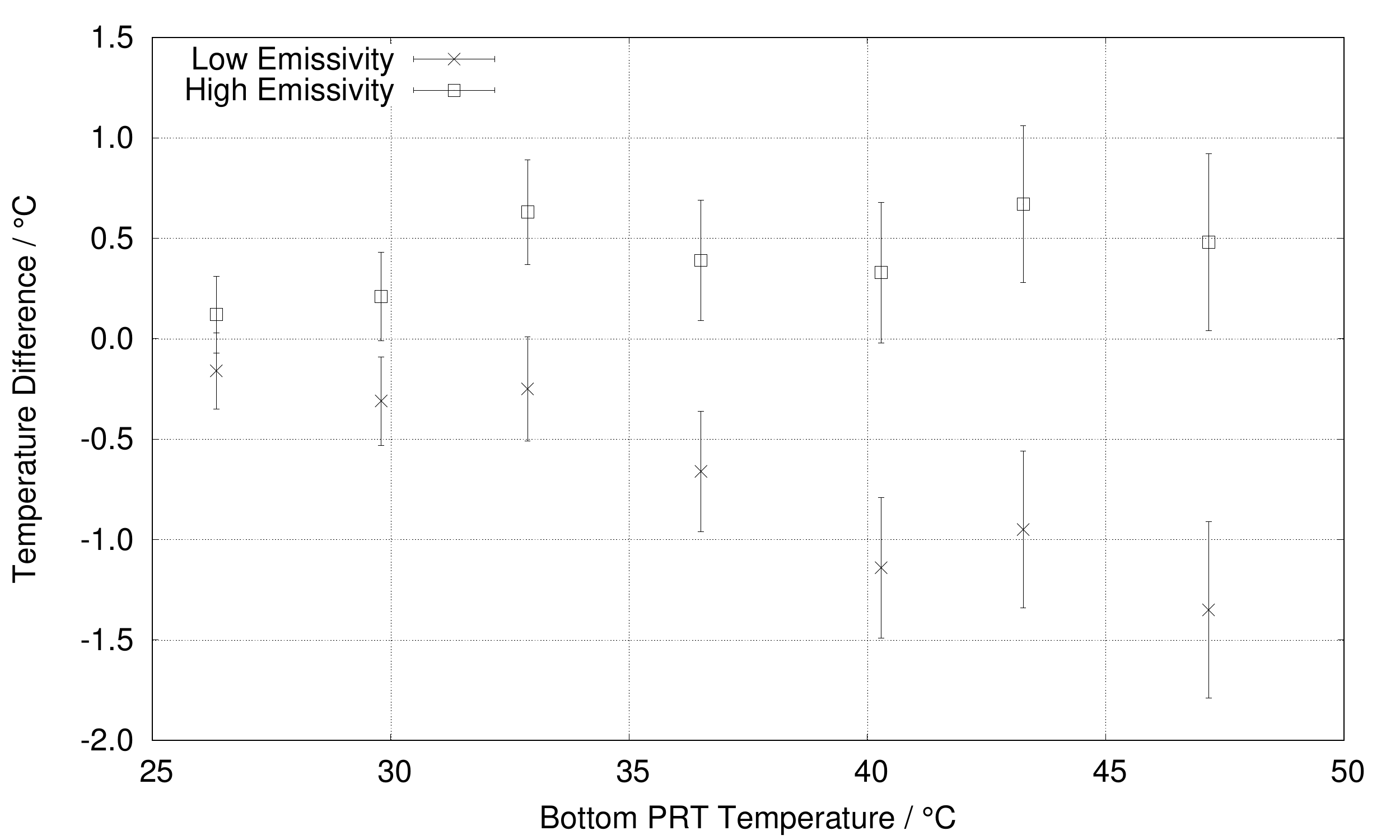}
\caption{From the MWIR measurements at the two angles of incidence, the difference in apparent temperature between the two angles (\(T\)\textsubscript{\SI{0}{\degree}}\( - T\)\textsubscript{\SI{-50}{\degree}}), for each coating, is depicted. The high emissivity vents appear to be approximately independent of temperature, whereas the low emissivity vents display a temperature dependence. The enhancement of emissivity due to the coating is small.}
\label{fig:results_thermal_imaging_angular_difference}
\end{figure}

The MWIR imager measurements are shown in Figure~\ref{fig:results_thermal_imaging_infratec}. The uncertainty in measurement is lower than that for the LWIR imager, and there is a distinct difference between the two angles of incidence which becomes more apparent as the temperature increases. Figure~\ref{fig:results_thermal_imaging_angular_difference} shows the temperature difference between the two angular positions as measured by the MWIR thermal imager. The high emissivity vents display zero temperature dependence, in addition the low emissivity vents demonstrate an emissivity enhancement at shallow angles.

\section{\bf Uncertainty Budget} \label{sec:uncertainty}

In this section an analysis of the uncertainty components associated with the thermometry techniques used is discussed.

Contact thermometry:

\begin{itemize}
\item {\bf Tolerance} - the stated tolerance for a class A Pt100 as described in \cite{ref:iso751}.
\item {\bf Calibration} - the maximum difference between ice point checks before and after measurements was \SI{0.04}{\celsius}.
\item {\bf Ice point repeatability} - the repeatability of an ice point is known to be in the order of \SI{0.01}{\celsius}.
\item {\bf Stability} - the largest standard deviation for each PRT at each setpoint throughout the total measurement campaign, then averaging across each PRT at each setpoint gave values ranging from \SIrange{0.03}{0.06}{\celsius}.
\end{itemize}

The uncertainty budget in Table~\ref{tab:uncertainty_budget} for the PRT measurements is at the highest temperature setpoint. Taking the quadrature summation of the components leads to a final uncertainty of approximately \SI{0.31}{\celsius} (\(k=2\)).

Thermal imaging:

\begin{itemize}
\item {\bf Calibration} - using the UKAS accredited procedure for the calibration of thermal imagers, the calibration uncertainty of the two thermal imagers was measured to be \SI{0.90}{\celsius} and \SI{0.15}{\celsius} for the LWIR and MWIR imager.
\item {\bf Sensor stability} - the stability of the internal sensor temperature for each imager during the measurement campaign was measured; this component was negligible for the MWIR imager, and was \SI{0.12}{\celsius} for the LWIR imager.
\item {\bf Repeatability} - the maximum standard deviation of individual vents was measured to be from \SIrange{0.2}{0.9}{\celsius} for the LWIR and between \SI{0.01}{\celsius} and \SI{0.05}{\celsius} for the MWIR imager.
\item {\bf Non-uniform emissivity} - the uncertainty due to vent-to-vent emissivity variation was modelled to be between \SI{0.2}{\celsius} and \SI{0.5}{\celsius} for the LWIR imager, and from \SIrange{0.1}{0.4}{\celsius} for the MWIR imager and coated vents.
\end{itemize}

The uncertainty budgets in Table~\ref{tab:uncertainty_budget} for the thermal imagers are at the highest temperature setpoint. Taking the quadrature summation of the components produces a final uncertainty of approximately \SI{1.3}{\celsius} (\(k=2\)) and \SI{0.4}{\celsius} (\(k=2\)) for the LWIR and MWIR thermal imagers.

Phosphor thermometry:

\begin{itemize}
\item {\bf ITS-90 traceability and surface extrapolation} - a Monte Carlo simulation evaluated the uncertainty as \SI{0.42}{\celsius} (\(k=1\)).
\item {\bf Coating thickness} - the maximum estimated thermal gradient is \SI{0.02}{\celsius}.
\item {\bf Coating stability} - a thermal cycling test (up to \SI{50}{\celsius}) shows no evidence of any systematic drift with repeated temperature cycles.
\end{itemize}

The uncertainty budget in Table~\ref{tab:uncertainty_budget} for the phosphor thermometer is at the highest temperature setpoint. Taking the quadrature summation of the components results in a final uncertainty of approximately \SI{1.0}{\celsius} (\(k=2\)).

\begin{table*}[h]
\centering
\caption{Uncertainty budget for temperature measurements. This depicts the uncertainty budget for the highest temperature setpoint. For measurements at lower temperatures, these values will have been smaller.}
\begin{tabular}{ !{\vrule width 2pt}M{4.00cm}|M{1.50cm}|M{1.00cm}|M{1.00cm}|M{2.00cm}!{\vrule width 2pt} }
\Xhline{2pt}
Source of Uncertainty & Estimate / \SI{}{\celsius} & Dist. & Divisor & Uncertainty (\(k=1\)) / \SI{}{\celsius} \\
\hline
\hline
\multicolumn{5}{!{\vrule width 2pt}c!{\vrule width 2pt}}{Contact thermometer} \\
\hline
Tolerance & 0.24 & R & \(\sqrt{3}\) & 0.14 \\
Calibration & 0.04 & R & \(\sqrt{3}\) & 0.03 \\ 
Ice point repeatability & 0.01 & R & \(\sqrt{3}\) & 0.01 \\ 
Stability & 0.06 & N & 1 & 0.06 \\ 
\hline
\multicolumn{4}{!{\vrule width 2pt}c}{Expanded uncertainty (\(k=2\))} & 0.31 \\
\hline
\hline
\multicolumn{5}{!{\vrule width 2pt}c!{\vrule width 2pt}}{LWIR thermal imager} \\
\hline
Calibration & 0.90 & N & 2 & 0.45 \\
Sensor stability & 0.12 & N & 1 & 0.12 \\ 
Repeatability & 0.38 & N & 1 & 0.38 \\ 
Non-uniform emissivity & 0.49 & N & 2 & 0.24 \\ 
\hline
\multicolumn{4}{!{\vrule width 2pt}c}{Expanded uncertainty (\(k=2\))} & 1.30 \\
\hline
\hline
\multicolumn{5}{!{\vrule width 2pt}c!{\vrule width 2pt}}{MWIR thermal imager} \\
\hline
Calibration & 0.15 & N & 2 & 0.08 \\
Sensor stability & 0.00 & N & 1 & 0.00 \\ 
Repeatability & 0.05 & N & 1 & 0.05 \\ 
Non-uniform emissivity & 0.40 & N & 2 & 0.20 \\ 
\hline
\multicolumn{4}{!{\vrule width 2pt}c}{Expanded uncertainty (\(k=2\))} & 0.44 \\
\hline
\hline
\multicolumn{5}{!{\vrule width 2pt}c!{\vrule width 2pt}}{Phosphor thermometer} \\
\hline
Standard uncertainty & 0.24 & N & 1 & 0.24 \\
ITS-90 calibration and surface extrapolation & 0.42 & N & 1 & 0.42 \\ 
Coating thickness variation & 0.02 & R & \(\sqrt{3}\) & 0.01 \\ 
Coating stability & 0.00 & R & \(\sqrt{3}\) & 0.00 \\ 
\hline
\multicolumn{4}{!{\vrule width 2pt}c}{Expanded uncertainty (\(k=2\))} & 0.97 \\
\Xhline{2pt}
\end{tabular}
\label{tab:uncertainty_budget}
\end{table*}


\section{\bf Discussion} \label{sec:discussion}
The objective of this study was to identify whether the {\em internal container temperature can be inferred from the temperature of the vent}. Using two thermometry techniques -- phosphor thermometry and thermal imaging -- the internal temperature was demonstrated to be proportional to the vent temperature as measured by each instrument; this is clear from Figure~\ref{fig:results_phosphor_vs_thermal_imaging}. Figure~\ref{fig:results_phosphor_vs_thermal_imaging} shows clearly the agreement between the phosphor and LWIR thermometry approaches within the measurement uncertainty; however there is a growing disparity between the two at higher temperatures. The correlation between vent temperature and PRT temperature is not one-to-one. However it is linear and given suitable characterisation both instruments could provide robust indication of the internal bulk temperature. The emissivity of the uncoated vent was estimated to be close to \num{0.9}, so the apparent temperatures measured by the LWIR imager were lower than that from the phosphor thermometer.

From the initial LWIR measurements, it was shown that within the measurement uncertainty there is no observable angular dependence of the low emissivity vent temperatures (Figure~\ref{fig:results_tau2_angular_low}). However an improved capability may show the angular (non-lambertian) variation of emissivity (bi-direction reflectance distribution function \cite{ref:brdf}). However this appears to depict the competing effects of geometrical enhancement of the vent and the specular behaviour of the vent surface.

\begin{figure}[h]
\centering
\includegraphics[width=0.45\textwidth,keepaspectratio]{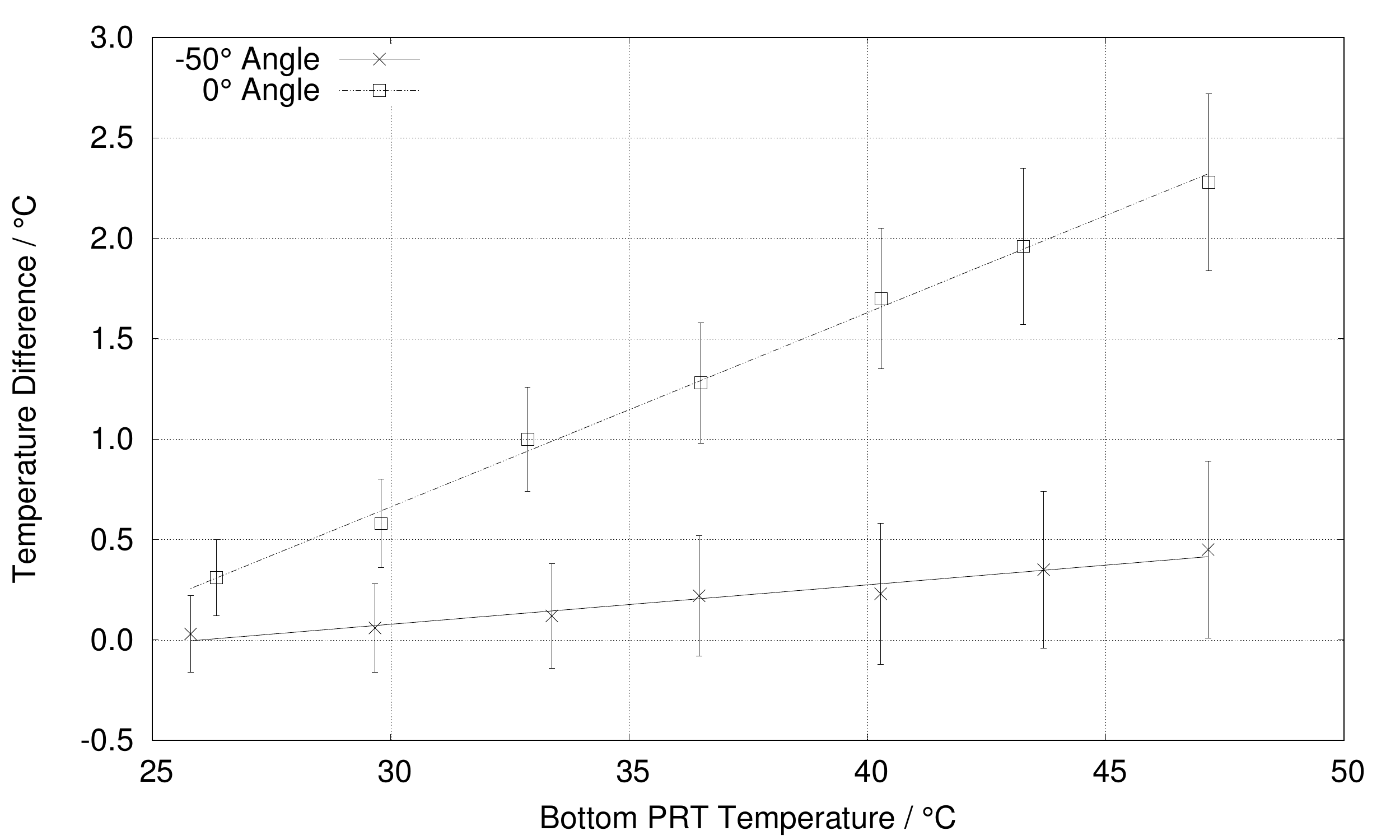}
\caption{For each setpoint measured by the MWIR, the difference between the temperatures measured at the high emissivity and low emissivity vents is plotted (\(T_{high \varepsilon} - T_{low \varepsilon}\)). The divergent behaviour indicates that at the \SI{-50}{\degree} position the geometrical enhancement of emissivity brings the two effective emissivities up to similar values.}
\label{fig:disc_infratec_angular_emissivity_difference}
\end{figure}

For the lower uncertainty MWIR measurements, a clear difference between the high emissivity and low emissivity vents at the two angles of incidence is apparent (Figure~\ref{fig:disc_infratec_angular_emissivity_difference}). The behaviour indicates that at the \SI{-50}{\degree} position the geometrical enhancement of emissivity brings the coated and uncoated effective emissivities up to similar values, indicated by the small temperature differences. This suggests that it is practical, and even preferred, to measure the vent temperature using a thermal imager at a shallower angle from the plane of the container lid in this case. 

\section{\bf Conclusion} \label{sec:conclusion}
The data presented in this report demonstrate the relationship between the temperature within an ILW container and the surface temperature as measured by a phosphor thermometer and two different wavelength thermal imagers. Both thermometry techniques displayed a similar linear correlation to the internal temperature. The angle of incidence for thermal imagers was shown to have a minor effect on the correlation, however it did highlight the geometrical enhancement of the vent geometry and demonstrated the lower measurement uncertainty for shallow angle measurements.

The largest contributor to the uncertainty of the thermal imaging measurement arises from the unknown surface emissivity of the vent. Providing the emissivity of the vent could be measured, it would be possible to reduce this component and so reduce the overall temperature measurement uncertainty.

Both phosphor and thermal imaging techniques successfully measured the correlation between the internal temperature of an ILW container and that of the external vent. It appears possible with appropriate engineering for these approaches to provide traceable temperature measurement for ILW vents and possibly containers.


\section{\bf Acknowledgements} \label{sec:acknowledgements}
This research was funded through a commercial contract between NPL and Sellafield; the authors would like to thank Sellafield for this opportunity.


\bibliographystyle{unsrt}
\bibliography{thermometry_of_ilw_containers}

\end{document}